\documentstyle[epsf,12pt]{article}
\begin{document}
\title{Searching for cold spots in multipion systems
}

\author{A.Bialas and K.Zalewski\thanks{also at the Institute of Nuclear
Physics, Cracow}\\ M.Smoluchowski Institute of Physics
\\Jagellonian
University, Cracow\thanks{Address: Reymonta 4, 30-059 Krakow, Poland;
e-mail:bialas@thp4.if.uj.edu.pl, Zalewski@chall.ifj.edu.pl}}
\maketitle

\begin{abstract}
Local fluctuations of pion density in momentum space may lead to
 Bose-Einstein
condensation. Conditions for this phenomenon to occur in high-energy collisions
and possibilities of its experimental investigation are discussed.
\end{abstract}

{\bf 1.} The possibility of Bose-Einstein condensation in a dense multipion system
was
first indicated in the pioneering work by Pratt \cite{pr1}. Recently,
the general solution of the problem has been obtained for gaussian
distributions \cite{zi1}  and also in the general framework of
uncorrelated production \cite{bz1,bz2,bz3}. In the present note, using the
results of
\cite{zi1}-\cite{bz3} we
discuss the conditions for creation of such a  pion condensate in
high-energy collisions and  the possibilities of its experimental
discovery.

The main conclusion of this investigation is that, although it seems
rather unlikely that the condensate may include all pions produced in the
collision, the possibility of its creation in a limited region of
phase-space is not excluded and thus worth experimental investigation.
The specific characteristics of the condensate which may help in its
experimental identification are

(i) rather small relative momenta of the pions ($\Delta p < 100$ MeV);

(ii) multiplicity distributions showing much stronger fluctuations than
those expected from the Poisson distribution;

(iii) large fluctuations in the charged/neutral ratio.

The first property requires  the "temperature" of the condensate to be
much smaller than that of its environement (characterized by the
average transverse momentum of about 350 MeV). Thus the search for the
condensate is in fact a search for "cold spots" in the pion system.

The actual existence of such "cold spots" is supported by the observation of
"intermittency" \cite{bp1}  showing that  large
non-poissonian fluctuations in very small momentum intervals do indeed exist
 and, moreover, that these fluctuations are
strongly related to the Bose-Einstein interference \cite{ddk}.

We thus propose to undertake a systematic search for the "cold spots" in
multiparticle production and, once they are identified, to investigate
their specific properties. We believe that such a program is feasible
and can provide interesting information on the structure of the systems
created in high-energy collisions.

{\bf 2.} The study of a general system of identical pions, which exibits
only correlations due to the quantum interference \cite{bz1,bz2,bz3} allowed
to formulate the necessary and sufficient condition at which
Bose-Einstein condensation takes place. Denoting by $\nu$ the assumed
average
multiplicity of pions before the quantum interference effects are taken
into account, the multiplicity distribution with quantum interference
included is described by the generating function of the form
\begin{equation}
\Phi(z)= \prod_m \frac{1-\nu\lambda_m}{1-\nu\lambda_mz},   \label{1a}
\end{equation}
where $\lambda_m$ are eigenvalues of the single pion density matrix,
$\rho^{(0)}(q,q')$ satisfying
 the normalization condition
\begin{equation}
\sum_m \lambda_m = 1.  \label{1b}
\end{equation}
 It
is clear from this formula that the distribution becomes singular when
\begin{equation}
\nu \lambda_0 \rightarrow 1,   \label{1}
\end{equation}
where $\lambda_0$ is the largest eigenvalue. This is precisely the point of condensation:
the average multiplicity tends to infinity and, moreover, almost all
particles occupy a single quantum state (corresponding to eigenvalue
$\lambda_0$) \cite{bz2}.

It is also seen from (\ref{1a}) that in the limit (\ref{1}) the first
factor in the product dominates and the distribution aproaches the
geometrical one
\begin{equation}
P(n) \rightarrow (1-\nu\lambda_0)[1,\nu\lambda_0, (\nu\lambda_0)^2,....]
\label{2}
\end{equation}
with the average
\begin{equation}
<n> \rightarrow \frac {\nu\lambda_0}{1-\nu\lambda_0}.  \label{3}
\end{equation}

Thus we conclude that close to the condensation point the multiplicity
distribution becomes very broad and differs drastically from the
original Poisson distribution of independently produced pions. This
broad distribution implies of course that the fluctuations in the
observed particle number must be very large. It was realized already by
Pratt \cite{pr1} (see also the recent discussion in \cite{sh1})
that this can be a good signal for observing this
phenomenon.

However, since the limit (\ref{1}) requires -- strictly speaking -- infinite
average multiplicity, and thus can never be reached in practice, it is
important to investigate what is the chance to observe this new regime
in real experimental conditions, i.e. at finite multiplicity. It is thus necessary
 to discuss the approach to the condensation limit.

To quantify this problem, it is convenient to consider the cumulants of
the distribution, which are easily derived from (\ref{1a}) \cite{bz1}
 \begin{equation}
K_p= (p-1)! \sum_m\left(\frac{\lambda_m \nu}{1-\lambda_m \nu}\right)^p.
\label{4}
 \end{equation}
In the limit (\ref{1}) we have
 \begin{equation}
\frac{K_p}{<n>^p} \rightarrow (p-1)!,   \label{5}
\end{equation}
whereas for the Poisson distribution all $K_p$ vanish.

It is now fairly clear that the approximation (\ref{2}),(\ref{5}) can
 be reasonable at finite multiplicities only
if the difference between the largest and the second largest
eigenvalue is not too small. In view of the normalization condition (\ref{1b}),
the sufficient condition for this is that $\lambda_0$ is close enough  to $1$.

Even if several terms contribute substantially to (\ref{4}), and thus
the approximation (\ref{5}) is not adequate, one may still have fairly
strong deviations from the original Poisson distribution
describing uncorrelated emission. Such deviations may serve as an
indicator of the approach to the condensation point. It is therefore
interesting to discuss them in more detail.

{\bf 3.} To estimate these effects of finite multiplicity, we have calculated the
multiplicity distribution following from the generating function
(\ref{1a}) in the case of a Gaussian density matrix of the form
 \begin{equation}
\rho^{(0)}(q,q') = \frac{1}{\sqrt{(2\pi\Delta)^3}}
\exp\left(-\frac{(\vec{q}^+)^2}{2\Delta^2}\right)
\exp\left(-\frac{R^2(\vec{q}^-)^2}{2}\right),   \label{6}
\end{equation}
where
\begin{equation}
\vec{q}^+=\frac{\vec{q}+\vec{q}'}{2};\;\;\;\; \vec{q}^- =\vec{q}-\vec{q}'.
\label{7}
\end{equation}
It is not difficult to verify that
\begin{equation}
<(\vec{q})^2> = 3\Delta^2;\;\;\;\; <(\vec{r})^2> = 3 R^2,   \label{8}
\end{equation}
which explains the physical meaning of these parameters. It is also not
very difficult to obtain the formula for $\lambda_0$ \cite{bz1}
\begin{equation}
\lambda_0= \left(\frac{2}{1+2R\Delta}\right)^3.      \label{9}
\end{equation}

In Fig. 1 the multiplicity distributions are  plotted for $<n>=3$ and
several values of $\lambda_0$. One sees radical deviations from
the Poisson distribution even for fairly small $\lambda_0$. The distributions
obtained
are much broader and extend to rather large multiplicities.
 We conclude that, as soon as $R\Delta$ becomes close to $.75$ (or smaller),
one may expect strong (and thus perhaps observable)
 effects on multiplicity distributions.  We have chosen a rather small
average multiplicity of $3$ identical pions to emphasize that we would like to consider
phenomena local in phase-space\footnote{One should remember, however,
that this means 9 pions on the average}.
 For larger multiplicities the effects
are  stronger, because one is closer to the condensation point. One
should keep in mind, however, that these results ignore the
energy-momentum conservation which tends to cut large
multiplicities and thus to reduce the deviations from the Poisson
distribution.

\begin{figure}
\epsfxsize 6cm
\begin{center}
~\epsfbox{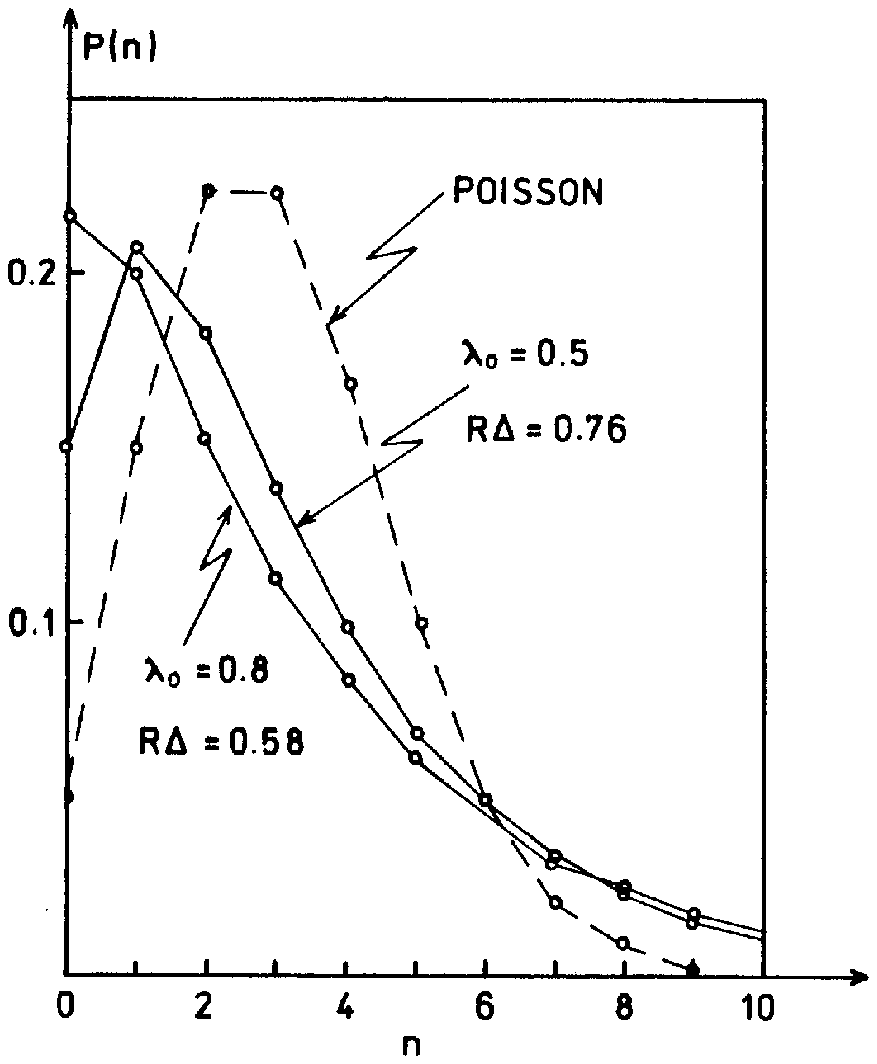}
\end{center}
{\bf Figure 1.} {\it Multiplcity distributions of identical bosons for $\langle
n \rangle = 3$.}
\end{figure}

{\bf 4.} It is seen from (\ref{8}) that the condition $R\Delta \leq .75$ implies
\begin{equation}
\sqrt{<(\vec{q})^2>}\sqrt{ <(\vec{r})^2>} \leq   \frac94 .   \label{10}
\end{equation}
Let us first discuss if a "standard" system created in a collision at
high energy may satisfy this condition. Taking the average transverse
momentum of about  $350$ MeV as approximate measure of $\sqrt{<(\vec{q})^2>}$:
\begin{equation}
  <(q_t)^2>= \frac23 <(\vec{q})^2> \label{11}
\end{equation}
we obtain from (\ref{10})
\begin{equation}
\sqrt{ <(\vec{r})^2>} \leq \frac{\sqrt{27/8}}{350 MeV} \approx 1 fm.
\label{12}
\end{equation}
It seems unlikely that such a small region can  contain  more than few
pions at "freeze-out" and therefore the analysis cannot apply to the
total multiplicity of an event. Moreover, even if one considers a
fraction of the produced particles, the meaningful comparison between the
distributions shown in Fig. 1, can only be performed if multiplicities
significantly larger than average are not reduced by requirement of a
a certain maximal density of pions at freeze-out.

On the other hand, one also sees from (\ref{10}) that even for a fairly
large volume of the system, the effects may be strong, provided
$<(\vec{q})^2>$ is small enough. For example, for $\sqrt{<(\vec{q})^2>}$
smaller than, say, 100 MeV - strong deviations from the Poisson distribution
appear even for average radius $\sqrt{ <(\vec{r})^2>}$ as large as 5
fm. Of course, such a small relative momentum inside a group of pions is a rare
phenomenon. However, once such a "cold spot" appears, it should have a very unusual
multiplicity distribution of identical particles.

{\bf 5.} The following comments are in order.

(i) The rather broad distribution of {\it identical} pions implies also
very large fluctutations in the charged/neutral ratio. Such large
fluctuations were suggested some time ago \cite{bk1} as a possible
signal for the disoriented chiral condensate. It was also shown that the
DCC is characterised by very small relative
momenta of pions \cite{rw1}. We thus conclude that it may be fairly difficult to
distinguish between DCC and "cold spots" solely on the basis of  measurements
of charged/neutral ratio.

(ii) A clear difference between DCC and the BE effect we discuss here is
in the distributions of identical particles. Indeed, the identical pions
emitted from DCC are expected to have a distribution close to the Poisson
one, whereas, as discussed in the present paper, this in not the case for
"cold spots". Consequently,  the simultaneous
measurements of charged and neutral distributions gives a chance to  distinguish
between the two phenomena.

(iii)  A search for "cold spots" in multiparticle systems created in
high-energy collisions
can probably
be performed with the help of the existing cluster algorithms. In view
of the arguments presented here, it seems worthwhile to
undertake such a systematic search.

(iv) A particularly attractive place to look for "cold spots" is the
region of small transverse momentum, where an enhancement in the pion
density was observed, particularly in heavy ion collisions. It is
believed that the majority of those pions are coming from resonance
decays \cite{sh1}. We would like to point out, however, that the origin of
the pions is irrelevant for our argument, provided the region of their
emission does not exceed the limit given by (\ref{10}). It is therefore
certainly interesting to verify what is the multiplicity distribution of
pions in the region of the "low $p_t$ enhancement".

(v) It was suggested \cite{pr1,bz1} that the phenomenon of BE condensation
could be perhaps at the origin of the so-called "centauro" and
"anticentauro" events \cite{fu1}. However,
since the observed "centauro" events show a
rather large relative momentum (of the order of 1 GeV)
between the particles belonging to the group, the present analysis implies that
this cannot be the case. Thus another explanation of these exotic events is
needed.

\vspace{0.3cm}
{\bf Acknowledgements}
This investigation was supported in part by the KBN Grant No 2 P03B 086
14.

\end{document}